\documentclass[12pt]{article}

\usepackage{verbatim} 
\usepackage{amssymb}

\newcommand{\beq} {\begin{eqnarray}} 
\newcommand{\eeq} {\end{eqnarray}} 
\begin {document}
\begin{titlepage}
\begin{center}
\vskip 6mm
{\Large \textbf{ On representations   of a chiral alternative to vierbein }}
\vskip 8mm

\textbf{Karmadeva Maharana}
\vskip 4mm
 {Physics Department\\
  Utkal University, Bhubaneswar 751004, India   }\\
{\tt karmadev@iopb.res.in}\\[1mm]

\end{center}
\vskip .2in
\begin{center} {\bf ABSTRACT } \end{center}
\begin{quotation}\noindent

In an attempt to facilitate the construction of a quantum theory
of gravity, 't Hooft has considered
a chiral alternative to the vierbein field 
in general theory of relativity. These objects,  
$ f^a{}_{\mu\nu}$, behave like the
``cube root'' of the metric tensor. We try to construct specific 
representations of these tensors in terms of Dirac 
$\gamma$ matrices in Euclidean and 
Minkowski space and promote these to curved space through Penrose-Newman 
formalism. We conjecture that these  new objects, with physical 
significance, are the analog 
of Killing-Yano tensors.
\end{quotation}
\vfill
\end{titlepage}
\eject

\section{Introduction}
A consistent and complete quantum formulation of gravity is yet to be
 realised. Leaving aside renormalization, just quantizing the theory
brings in many problems. Even for pure gravity, besides the problems 
arising at the 
Planck scale, 
many conceptual and calculational
 difficulties are encountered in quantising gravity. This is 
because diffeomorphism 
invariance is  to be applied to space-time itself and the background 
independence of the action \cite{rov}. 
 To overcome these, 
 there have been  attempts through different approaches and a 
variety of  methods have been used  to tackle each of the problems 
encountered.
 These have resulted in
 varying amount of successes  in addressing  problems dealing with 
particular aspects \cite{carlip}. 
The problems with quantizing even pure gravity have been analyzed 
by Nicolai and Peeters \cite{Nicolai}.

The very existence of matter fields, and in particular the fermions, 
necessitates  their 
incorporation in a complete  quantisation programme. Unlike
tensors, there is no spinor
representation for the diffeomorphism invariance corresponding to 
the $ GL(4)$
general coordinate transformation of general relativity
\cite{wein}.
   The only way to 
incorporate spinors in general relativity is through the introduction 
of vierbeins or tetrads.
A tetrad behaves like the ``square root'' of the metric tensor
$ g_{\mu \nu}$. Using this it has been possible to incorporate
the spinor, which is  `` square root'' of a vector or tensor of rank one,
into general relativity.
The construction of both fermions and bosons out 
of spinors and their various combinations has been exploited through 
this approach.
 Tetrads were introduced by Weyl \cite{weyl} and are useful in some
 cases 
in calculating  Riemann curvature and 
Ricci tensors. Use of tetrads simplifies calculations and the 
results expressed 
in terms of the tetrads are in more easily understandable 
form \cite{lan2}.  By the choice of tetrads, when tensors are  written 
via  tetrad components, the tensors 
can also be  expressed in a  coordinate independent fashion and their 
algebraic properties
  become more transparent and the tensor components can be
 simplified \cite{step}.
 The variation of the action with respect to tetrads or their higher 
dimensional incarnations,  and the spin connections, 
gives the Einstein field equations. Such an action is the basis 
 for  quantum formulation of gravity in string theory as well as in 
other covariant formulations, and also in the canonical approach.
The application of tetrad formalism has been further explored in 
\cite{ch}.
In the canonical approach to quantum gravity these appear ultimately 
through Ashtekar 
variables \cite{Ashtekar}.

Another class of objects, closely connected to 
the vierbein, are the self-dual or anti-self-dual
solutions to the Yang-Mills equations of the gauge theories
in the form of monopoles, instantons etc. 

Pure quantum gravity was found to be one-loop finite 
on shell by 't Hooft and Veltman. However, adding scalar matter 
field made the
theory nonrenormalizable \cite{'t Hooft2}.  Goroff and Sagnotti found that
the conventional quantum field theoretic methods, that has been reasonably 
successful in describing the electro-weak and strong interactions at energy 
scales presently accessible, fail in the case of gravity \cite{sagnotti}.
In the context of   string theory formulation, the programme to 
obtain all the interactions in a unique manner
 appears to have too many
choices due to the landscape of the ground state \cite{Douglas}.  

It would be interesting to have a theory where the SU(3) symmetry 
of QCD comes out naturally in a manner similar to the Dirac's 
relativistic 
theory of electrons (fermions) where the spin, magnetic 
moments and other physical 
consequences come out automatically.
In the chiral alternative to vierbein \cite{tH}, there is a rare natural 
appearance of
an SU(3) symmetry, although it does not seem to be related 
to Quantum Chromodynamics.
However, it is tempting to speculate that the elementary fields
could have various symmetries under $U(1)$, $ SU(2)$ and $SU(3)$ 
representations,
somehow  arising naturally from the chiral alternative.
A chiral alternative to the vierbein was formulated by 't Hooft
as an object with interesting characteristics
that reflect some properties of the dual Yang-Mills solutions.
Another  motivation for looking at
the chiral alternative to the vierbein is to analyse
the "cube root of the metric tensor" so as to obtain
some objects analogous to the Killing-Yano tensors.
One would expect these to lead to symmetries, 
conservation laws and separability of the equation \cite{Page}.


The canonical approach was initiated by trying to put
the Einstein equation into a  Hamiltonian 
form \cite{Bergmann,Dirac,Schild}.
As is well known, even in some classical Lagrangians there appears 
problems in finding the canonically
conjugate momenta, such  as the gauge freedom in electrodynamics 
and massive vector particles \cite{wein2},
 and in case of gravity these are further exacerbated
 by complications
due to the full diffeomorphism invariance. One approach in formulating 
quantum  gravity is through respecting this invariance, 
which is a non-perturbative background independent approach. 
 This analogue of gauge invariance of electrodynamics
is a  huge freedom and to  tackle this the  formalism of 
constraints was developed by Dirac in his attempt to quantize
gravity.
 In canonical approach, the  diffeomorphism invariance
 for the spatial part is manifestly kept. The $D +1$ dimensional
 manifold, $M$ is assumed to have the special topology
of $ M = {\mathbb{ R}} \times \sigma $, where $\sigma $ is a fixed, 
$D$ dimensional (where $D$ is taken to be three), compact manifold 
without boundary.  The first and 
 the second fundamental forms, pulled back to  $\sigma $,
are the `` spatial'' tensors used as the basic ingredients of the 
 action and the Legendre transformed Hamiltonian. These relate to the 
curvature and Christoffel connection of the four dimensional space
 via the Gauss-Codacci equation and also serve as the basic dynamical 
variables along with the lapse function and shift vector. 
This results in a singular Lagrangian. The vanishing of conjugate 
momenta for the lapse and shift
 are the primary Dirac constraints 
  and the consistency of the equations of motion leads to the secondary 
 constraints which are 
 the diffeomorphism constraint and the Hamiltonian constraint.
This results in making the Hamiltonian constrained to vanish.
 Treating the lapse function and shift vector as Lagrange multipliers, one 
arrives at the Arnowitt-Deser-Misner action \cite{Arn}. Analysis 
of the $ADM$ action shows that 
the spacetime diffeomorphisms 
generated by the Lie algebra of a symmetry group is implemented
in the canonical framework. The 
basic variables, namely, the first  fundamental form \cite{Geroch} 
and the corresponding
 momentum  
are not observables  as they are not gauge invariant.
The classical constraint functions depend nonpolynomially (also 
nonanalytically) on the metric which is identified as the first 
fundamental form. The curvature scalar depends on the inverse of this 
metric. Since the products of the metric and  the conjugate 
momentum also appear in the constraint equations, this leads to 
difficulties in quantizing the theory. 
By breaking up the above D dimensional metric in terms of an 
$su(2)$-valued one form, which are
D-bein fields on $\sigma $ and by introducing the Sen variables
\cite{Sen}, Ashtekar
with  a subtle choice of the spin connection
variables  was able to modify the theory
in such a way so as to get the constraints in a polynomial 
form \cite{Ashtekar}. It has been shown that 
general relativity arises out of a connection dynamics 
which parallel transports
the chiral fermions. Here, triads (analogue of tetrads in three 
dimensions) and the connections are the conjugate variables
and  the constraint equations of genral relativity looks similar to
 those of Yang-Mills theories. 
In early works Einstein and Schroedinger had used affine connections as 
basic variables
and simplification occurs for the chiral fermions \cite{Au}.
Further generalisation by Immirzi, Barbero, Rovelli, Smolin and 
others resulted in the loop
formulation in terms of spin network \cite{rov}.
However, a satisfactory quantisation of gravity has  yet to be 
achieved in any formulation.
Hence, alternate approaches are needed to be explored. 

One such approach is the alternate to chiral vierbein by 't Hooft. 
Though the formulated Lagrangian is not renormalizable, the 
chiral alternative has many interesting features, such as, 
emergence of new symmetry, self(anti)-duality etc. In this paper, 
we have tried to construct the representations of the chiral 
alternatives in terms of Dirac $\gamma$ matrices in curved space. 
In section 2, we briefly introduce the chiral alternative. In 
section 3, we work out the representations in flat Minkowski 
as well as Euclidean space and we give a generalization 
to curved space. In section 4, we conjecture a generalization 
of Killing-Yano tensor and discuss the future prospective. 
 
\section{ A chiral alternative}
 Vier(l)beins and chiral fermions  play a fundamental role
in most formulations of quantum gravity, such as, covariant, 
canonical, loop and string formulations. 
In an attempt to better understand quantum gravity, 
't Hooft has introduced an alternate new interesting object that 
behaves like the 
``cube root'' of the metric tensor, instead of the vierbein.
 Following 't Hooft,
 we briefly review the role of different fields leading to the
chiral alternative and in setting up the basic formalism \cite{tH}.

For introduction of Dirac field to general relativity, as well as an
alternative  to the metric tensor as the fundamental variable in the
covariant formalism, it is useful to
introduce the ``square root'' of the metric tensor $g_{\mu\nu} $, the
vierbein  field ${e^a}_\mu $,
\begin{eqnarray}
  g_{\mu\nu} = {e^a}_\mu \;{e^a}_\nu . \label{eq:g}
\end{eqnarray}
Here $\mu $, $\nu $ are four vector indices, $a$ represents 
``internal'' indices. 
$ {e^a}_\mu $ has 16 components but $ g_{\mu\nu} $ has 10 independent
ones.
So 6 degrees of freedom should reside in internal $O(3,1)$. 
For the covariant derivative of the vierbein, 
 a connection field $ {{A^{ab}}_\mu}$ corresponding to this local symmetry is
  introduced. The metric being covariantly constant, we expect the
vierbein to be so and require 
\begin{eqnarray}
  D_\mu\ {e^a}_\nu =   {\partial}_\mu \; {e^a}_\nu -
  {{\Gamma}^{\lambda}}_{\mu\nu}\; {e^a}_\lambda + {A^{ab}}_\mu \; {e^b}_\nu
  = 0 . \label{eq:A}
\end{eqnarray}
 ${{\Gamma}^{\lambda}}_{\mu\nu}$ has 40, and ${A^{ab}}_\mu $ has 24 degrees
 of freedom, respectively. \\
${R^{\alpha}}_{\beta\mu\nu} $ can be expressed in terms of $\Gamma $
 and $A$ fields as
\begin{eqnarray}
 \left[D_\mu , D_\nu \right]  {e^a}_\beta = -
 {R^{\alpha}}_{\beta\mu\nu} \; {e^a}_\alpha + {F^{ab}}_{\mu\nu}\;
 {e^b}_\beta  = 0,
\end{eqnarray}
where
\begin{eqnarray}
 {F^{ab}}_{\mu\nu} = {\partial}_\mu {A^{ab}}_\nu - {\partial}_\nu
 {A^{ab}}_\mu + {\left[ A_\mu , A_\nu \right]}^{ab}
\eeq
implying
\begin{eqnarray}
 {R^{ab}}_{\mu\nu} =  {F^{ab}}_{\mu\nu}.
\eeq
In terms of $  {F^{ab}}_{\mu\nu}  $ and  $ {e^\mu}_a  $ the
Lagrangian leading to Einstein equation becomes
\begin{eqnarray}
\mathcal{L} = \sqrt{g}\; R = det(e) \; {F^{ab}}_{\mu\nu} \; {e^\mu}_a \;
 {e^\nu}_b
\eeq
So a simple polynomial Lagrangian is obtained, 
with independent variations of $ {A^{ab}}_\mu $ and $ {e^a}_\mu $ 
giving the Einstein equation.

Using the 't Hooft  invariant self-dual tensor, introduced in the monopole
like
solutions\cite{tH2},
\begin{eqnarray}
{{\eta}^a}_{\mu\nu} = - {{\eta}^a}_{\nu\mu} = {\epsilon}_{a\mu\nu} 
 + {\delta}_{a\mu}\; {\delta}_{\nu 4}  - {\delta}_{a\nu}\; {\delta}_{\mu 4}, 
\end{eqnarray}
where $ a = 1,2 $ or $3$, $\epsilon $ is 3-dim Levi-Civita symbol,
't Hooft has introduced a field $ {e^a}_{\mu\nu}$ in curved
space-time that takes the values ${\eta^a}_{\mu\nu}$ in a locally flat
coordinate frame. It satisfies, 
\begin{eqnarray}
{\epsilon}_{abc} \; {{e}^a}_{\mu\nu} \; {e^b}_{\kappa\lambda}\;
{e^c}_{\rho\sigma} \; {\epsilon}^{\mu\nu\kappa\rho } = 24 \sqrt{g}\;
g_{\lambda\rho} .\label{eq:B}
\end{eqnarray}

This is invariant under any transformation of the form
\begin{eqnarray}
{e^a}_{\mu\nu} \Rightarrow {S^a}_b \; {e^b}_{\mu\nu} 
\end{eqnarray}
where, ${S^a}_b \in SL(3)$, for the Euclidean space and 
 $det\: S = 1$. 
Analogous to vierbein field, one introduces an $SL(3)$ connection field
${A^a}_{b\mu}$ by demanding
\begin{eqnarray}
  D_\mu \ {e^a}_{\alpha\beta} =   {\partial}_\mu  {e^a}_{\alpha\beta} -
  {{\Gamma}^{\lambda}}_{\mu\alpha} \; {e^a}_{\lambda\beta}  
  - {{\Gamma}^{\lambda}}_{\mu\beta} \; {e^a}_{\alpha\lambda} + 
  {A^{a}}_{b\mu}\; {e^b}_{\alpha\beta}
  = 0 . \label{eq:cA}
\end{eqnarray}
This leads to \cite{tH}, 
\begin{eqnarray}
  {A^{a}}_{a\mu} = 0
\end{eqnarray}
 and a bilinear expression in $ {e^a}_{\alpha\beta}$ :  
\begin{eqnarray}
  K^{ab} = \frac{1}{8}\; {\epsilon}^{\alpha\beta\mu\nu} \;
  {e^a}_{\alpha\beta} \; {e^b}_{\mu\nu} 
\end{eqnarray}
which has an inverse $K_{ab}$.

By redefining 
\begin{eqnarray}
 {e^a}_{\mu\nu} \; (detK)^{-\frac{1}{9} } = {f^a}_{\mu\nu} 
\end{eqnarray}
equation ({\ref{eq:B}}) can be rewritten as,
\begin{eqnarray}
{\epsilon}_{abc}\; {{f}^a}_{\mu\nu} {f^b}_{\kappa\lambda}
{f^c}_{\rho\sigma}  {\epsilon}^{\mu\nu\kappa\rho } = 24\;
g_{\lambda\rho} .
\end{eqnarray}
where, $ {f^a}_{\mu\nu}$ is the chiral alternative. The   
Lagrangian now becomes
\begin{eqnarray}
  \mathcal{L} = \frac{1}{32} \; {\epsilon}^{\kappa\lambda\rho\sigma } \;
  {f^c}_{\kappa\lambda} \; {f^b}_{\rho\sigma}\; {F^a}_{c\mu\nu}\;
  {\epsilon}_{abd} \;{f^d}_{\alpha\beta} \;{\epsilon}^{\mu\nu\alpha\beta}
\end{eqnarray}
with 
\begin{eqnarray}
  {F^{a}}_{b\mu\nu} = {\partial}_\mu {A^{a}}_{b\nu} - {\partial}_\nu
 {A^{a}}_{b\mu} + {{\left[ A_\mu , A_\nu \right]}^{a}}_{b}  \nonumber \\
  =
 {\frac{1}{2}}\; {\epsilon}_{abd} \;{{\eta}^d}_{\lambda\alpha}\;
 R_{\lambda\alpha\mu\nu} .
\end{eqnarray}
This gives a relation between $f$ and the metric $g$, i.e.,
for ``cube root'' of $g_{\kappa\tau}$,
\begin{eqnarray}
 {\epsilon}_{abc}\; {\epsilon}^{\mu\nu\lambda\rho}\; f^a{}_{\mu\nu} \;
 f^b{}_{\lambda\kappa} \; f^c{}_{\rho\tau}  = 24\;\;
  g_{\kappa\tau}  \label{eq:tH}
\end{eqnarray}
The imposition of the gauge condition and  the consequent 
constraint term  added to the Lagrangian  makes it
non-renormalizable.
In Minkowski space, putting a reality condition
\begin{eqnarray}
 {{\hat{f}}^a}_{\mu\nu} =  {\left({{{f}}^a}_{\mu\nu}\right)}^{\ast} 
\label{eq:su3}
\end{eqnarray}
converts the internal $SL(3)$ to $SU(3)$ symmetry.

\section{An attempt to obtain a representation in flat space and 
generalization to curved space}

Here, we try to construct  $ f^a{}_{\mu\nu}$ in terms 
of Dirac { $\gamma^\mu $}
matrices to get a relation related to ({\ref{eq:tH}}) for flat 
Euclidean space
and also for  Minkowski space. Then, as for the generalisation of Dirac
equation to curved space, we can promote 
the  $\gamma^\mu $ matrices, and hence  $ f^a{}_{\mu\nu}$  to
the curved space.

\subsection{Euclidean space }
The 't Hooft tensor, 
introduced in the  context of monopole-like solutions is given by, 
\begin{eqnarray}
  \eta^a{}_{\mu\nu} = \epsilon_{a\mu\nu} + g_{4\mu}\; g_{\nu a} -
  g_{4\nu}\; g_{\mu a}  
\end{eqnarray}
where, $(a = 1,2,3)$ and it has the property of being antisymmetric 
and self-dual.

If we express $ f^a{}_{\mu\nu}$ as
\begin{eqnarray}
 f^a{}_{\mu\nu} = \sigma^a \; \gamma_5 \; \sigma_{\mu\nu} \label{eq:fE}
 \end{eqnarray}
 with  $\gamma^\mu$ matrices in Pauli-Dirac representation \cite{sak},
\begin{eqnarray}
\sigma_{\mu\nu} = \frac{1}{2i} \left[\gamma_\mu ,\gamma_\nu \right],
\nonumber \\
\gamma_\mu \gamma_\nu + \gamma_\nu \gamma_\mu =
2\delta_{\mu\nu},\nonumber \\
\gamma_5 = \gamma_1 \gamma_2 \gamma_3 \gamma_4 = \frac{1}{4}
\epsilon_{\mu\nu\lambda\rho} \;\gamma_\mu \gamma_\nu \gamma_\lambda
\gamma_\rho ,\nonumber \\
 {\gamma_\mu}^\dag = \gamma_\mu
\end{eqnarray}
then  $ f^a{}_{\mu\nu}$ is antisymmetric in $\mu$ and $\nu $ and is
anti-self-dual modulo $\gamma_5$. Here $ f^a{}_{\mu\nu}$ appears as an
object in the direct product of two spaces, the internal space indices
$a,b,c$ and the 4-space Greek indices
$\mu, \nu \ldots $ etc.

For $\kappa = \tau $, one can show that, 
\begin{eqnarray}
\epsilon_{abc}\; \epsilon_{\mu\nu\lambda\rho}\; f^a{}_{\mu\nu}\;
f^b{}_{\lambda\kappa}\;f^c{}_{\rho\tau} \nonumber \\ 
  = \epsilon_{abc}\;
\epsilon_{\mu\nu\lambda\rho} \;\sigma^a \; \sigma^b \; \sigma^c \;\gamma_5
\gamma_\mu \gamma_\nu \; \gamma_5 \;\gamma_\lambda \gamma_\kappa\; \gamma_5\;
\gamma_\rho \gamma_\tau \nonumber \\
 = 6i\; I_2 \times 12\; I_4  
\end{eqnarray}
This expression must vanish for $\kappa \neq \tau$. 
To make the right hand side vanish for $\kappa \neq  \tau $, we may
take either the trace or the determinant of $ \epsilon_{abc} \;
\epsilon_{\mu\nu\lambda\rho}\; f^a{}_{\mu\nu}\;
f^b{}_{\lambda\kappa}\;f^c{}_{\rho\tau} $, so that 
\begin{eqnarray}
 tr \left(  \epsilon_{abc} \;
\epsilon_{\mu\nu\lambda\rho}\; f^a{}_{\mu\nu}\;
f^b{}_{\lambda\kappa}\;f^c{}_{\rho\tau} \right ) \propto g_{\kappa\tau}
\end{eqnarray}
Here, the expression on r.h.s. means value of that component. 
One can also use, 
\begin{eqnarray}
 det \left(  \epsilon_{abc}\; 
\epsilon_{\mu\nu\lambda\rho}\; f^a{}_{\mu\nu}\;
f^b{}_{\lambda\kappa}\;f^c{}_{\rho\tau} \right ) \propto g_{\kappa\tau}
\end{eqnarray}
where $g_{\kappa\tau} $ is the Euclidean flat space metric.

We can also get the general relation,
\begin{eqnarray}
 \epsilon_{abc}\; 
\epsilon_{\mu\nu\lambda\rho}\; f^a{}_{\mu\nu}
f^b{}_{\lambda\kappa}\;f^c{}_{\rho\tau} + ( \kappa \rightleftharpoons
\tau )  =         \nonumber  \\
 6i\; I_2 \times 12 \;I_4 \;g_{\kappa\tau} \label{eq:tHE}
\end{eqnarray}
So the relation ({\ref{eq:tH}}) is not produced exactly, but
equation ({\ref{eq:tH}}) satisfies equation ({\ref{eq:tHE}}) modulo
a constant and identity matrices. 
Since we have considered in the simplest case $ \sigma^a $
to be two dimensional, and the space to be a direct product space,
it would be more appropriate to call our construct as an attempt 
to  a realisation rather than
a representation. It is also intriguing that due to the presence of
$ \gamma$'s in equation 
({\ref{eq:fE}}), $  f^a{}_{\mu\nu}$   can act on spinors.

\subsection{Minkowski space }
In the Minkowskian case,
we again define 
\begin{eqnarray}
f^a{}_{\mu\nu} = \sigma^a \; \gamma_5\; \sigma_{\mu\nu}
\end{eqnarray}
which is antisymmetric in $\mu$ and $\nu$ and is anti-self-dual modulo
$i\gamma_5 $. Here we use the $\gamma_\mu $ matrices of Bjorken and
Drell \cite{bj}
\begin{eqnarray}
\sigma_{\mu\nu} = \frac{1}{2i} \left[\gamma^\mu ,\gamma^\nu \right],
\nonumber \\ 
\gamma_\mu \gamma_\nu + \gamma_\nu \gamma_\mu =
2 g_{\mu\nu},\nonumber \\
 g_{\mu\nu} = ( 1, -1 , -1 , -1 ) \nonumber \\
\gamma_5 = -i \gamma_0 \gamma_1 \gamma_2  \gamma_3 .
\end{eqnarray}

 We get, for $\kappa =\tau$ ,
\begin{eqnarray}
\epsilon_{abc}\; \epsilon_{\mu\nu\lambda\rho}\; f^a{}_{\mu\nu}
f^b{}_{\lambda\kappa}\;f^c{}_{\rho\tau}  =  6i\; I_2 \times (-i)\;12
\;I_4\;
g_{\kappa\tau} 
\end{eqnarray}
To make it vanish for $\kappa \neq \tau$, we consider like in the
Euclidean case and obtain, 
\begin{eqnarray}
\epsilon_{\mu\nu\lambda\rho} \;f^a{}_{\mu\nu}\;
f^b{}_{\lambda\kappa}\;f^c{}_{\rho\tau} + ( \kappa \rightleftharpoons
\tau )  = 144\; I_2 \times  I_4 \; g_{\kappa\tau}
\end{eqnarray}
or take the trace of the left hand side to obtain, 
\begin{eqnarray}
 tr \left(  \epsilon_{abc}\; 
\epsilon_{\mu\nu\lambda\rho}\; f^a{}_{\mu\nu}
f^b{}_{\lambda\kappa}\;f^c{}_{\rho\tau} \right ) \propto g_{\kappa\tau} .
\end{eqnarray}
We can also construct another object
\begin{eqnarray}
\breve{f}^a{}_{\mu\nu} = \sigma^a \;\left( g_{\mu\mu} \; g_{\nu\nu}
+ i \gamma_5 \right) \;\sigma_{\mu\nu}
\end{eqnarray}
which is antisymmetric in $\mu$ and $\nu$ and is anti-self-dual. Its properties
are similar to that of $ f^a{}_{\mu\nu}$.
However, promoting this to curved space will have problem as both the
left and right side would contain the metric tensor $g_{\mu\nu}$,
unless we write 
\begin{eqnarray}
\breve{f}^a{}_{ij} = \sigma^a\; \left( 1
+ i \gamma_5 \right)\; \sigma_{ij}, \nonumber\\
\breve{f}^a{}_{i0} = \sigma^a \;\left(- 1
+ i \gamma_5 \right)\; \sigma_{i0}.
\end{eqnarray}

\subsection {Curved space}
To go from flat Minkowski space to curved space-time we follow 
 Chandrasekhar \cite{ch} in making use of
Penrose-Newman formalism. For this, the constant 
Pauli matrices, { $\sigma_i$},
constituting the {$\gamma_{\mu}$} matrices are replaced by,
\begin{eqnarray}
\sigma^i{}_{A{B'}} = \frac{1}{\sqrt{2}} 
               \left| \begin{array}{cc}
                l^i         & m^i \\
                {\bar{m}}^i & n^i  \end{array} 
               \right|     
\end{eqnarray}
where {\boldmath ${{l}, {m}, {\bar{m}},{n} }$} are the null 
basis vectors \cite{ch} and $A$, $B'$ are the spinor indices. 
 This way we  obtain  $ f^a{}_{\mu\nu}$
for a curved space.

\section{Conclusion and outlook }

Killing tensors and Killing-Yano tensors give constants of geodesic
 motion that are in evolution. By enumeration of the constants in 
involution,
the geodesic motion becomes completely integrable. These are also 
related to the constants in the separation of the Hamilton-Jacobi and 
Klein-Gordon equations \cite{Page}.
The ``square root'' of a Killing tensor of order two, the 
Killing-Yano tensor encode the symmetries of the theory 
and is related to the quadratic 
first integrals of the geodesic motion, as well as, to the separability of 
the partial differential equation. \\
For matter coupled to gravity, where there is coupling to 
higher spin states, the acceleration
is expected to depend on higher powers of the four velocity
and this may give rise to new type of conserved quantities \cite{deWit}.
Symmetries of spinning particles in Schwarzschild and Kerr-Newman type
space times have been analysed by Gibbons, Rietdijk and van Holten
\cite{gib}. They found new nontrivial supersymmetry corresponding to
the Killing-Yano tensor for the black-hole space-time. This also plays an 
important role in solving the Dirac equation in these black-hole metrics.
The fermionic symmetries are generated by the square root of 
bosonic constants of motion other than the Hamiltonian.

The new object, the 't Hooft tensor 
 $ f^a{}_{\mu\nu}$ itself has many interesting properties. 
The use of twistors in producing anti-self-dual solutions of the Yang-Mills
equations  indicates these are related to 't Hooft tensor  $ f^a{}_{\mu\nu}$. 
We also note  that an internal $SU(3) $ transformation of 
$ f^a{}_{\mu\nu}$ appears for the Minkowski space. 
In another context, an $SU(3) $ arises
naturally  from the Runge-Lenz type of symmetry in harmonic oscillator
 potential
\cite{Schiff}, which is a consequence of a dynamical symmetry.
However, here the  $SU(3) $  is not the consequence of a potential and 
its origin is geometrical.
It is natural to expect conserved quantities arising out of the $SU(3) $
symmetry of equation ({\ref{eq:su3}}). 
It would be interesting to relate these to
the Killing-Yano type symmetries.

Therefore, one would like to conjecture that the ``cube
root'' $ f^a{}_{\mu\nu}$ would have analogous relation with the symmetries
and the  conserved quantities. This Killing-'t Hooft tensor 
$ f^a{}_{\mu\nu}$
would be a new mathematically interesting object to analyse.
Establishing such a connection is expected to lead to a better
 understanding of the structure of space-time.

\subsection*{Acknowledgements}
{ I am grateful to Professor G. 't Hooft for critical comments.
I wish to thank Professor B. de Wit and Professor G. 't Hooft   for the
kind hospitality at Spinoza Institute,
University of Utrecht, where part of this work was carried out.
Financial support from 
Nederlandse Organisatie voor Wetenschappelijk Onderzoek,  (NWO) is 
acknowledged with gratitude.
}


\begin{thebibliography}{99}
\bibitem{rov} C. Rovelli, Quantum Gravity, Cambridge University Press,
    (2004);"Loop Quantum Gravity", Living Rev. Relativity 1 (1998);
     T. Thiemann, "Lectures on Loop Quantum Gravity", Lect. Notes 
     Phys. 631: 41–135 (2003).

\bibitem{carlip} S. Carlip, gr-qc/0108040.

\bibitem{Nicolai} H. Nicolai and K. Peeters, Lect.Notes Phys.721:151-184
(2007).

\bibitem{wein} S. Weinberg, Quantum Theory of Fields, Vol 3,
                Cambridge University Press, (2000);
                Gravitation and Cosmology: Principles of the 
                general theory of relativity, John Wiley, (1972). 

\bibitem{weyl} H. Weyl, Z. Phys. {\bf{56}}, 330 (1929); 
                Phys. Rev. {\bf{77}},699 (1950) 

\bibitem{lan2} L. D. Landau and E. M. Lifshitz, The Classical Theory 
                of Fields (4th Edition), Elsevier, Amsterdam, 2005;\,
            T. W. B. Kibble, J.Math. Phys. {\bf{2}}, 212 (1961);
           J.Math. Phys. {\bf{4}}, 1433 (1963);
           S.Deser and C. J. Isham,
             Phys. Rev.D  {\bf{14}}, 2505, 1976.

\bibitem{step} H. Stephani, General Relativity,
        An introduction to the theory of the gravitational field,
       (Second edition), Cambridge University Press, Cambridge, (1990).

\bibitem{ch} S. Chandrasekhar, The Mathematical Theory of Black Holes,
             Oxford University Press, (1981).

\bibitem{Ashtekar} A. Ashtekar,
                  Phys. Rev. Lett. {\bf{57}}, 2244 (1986);
             P. G. Bergmann and G. J. Smith,
            Phys. Rev.D {\bf{43}}, 1157 (1991); Phys. Rev.D {                                 \bf{36}}, 1587 (1987).
          
               Phys. Rev.D {\bf{36}}, 1587 (1987).
           
\bibitem{'t Hooft2} G. 't Hooft and M. Veltman, Ann. Inst. Henri 
                   Poincar{\'{e}} Phys. Theor. {\bf {A20}}, 69 (1974).


\bibitem{sagnotti}M. H. Goroff and A. Sagnotti, Phys. Lett.
                        {\bf{B160}} (1985) 81;
                                Nucl. Phys.
                          { \bf{B266}} 709(1986);
                         A. E. M. van de Ven, Nucl. Phys. B378 (1992) 309.

\bibitem{Douglas} M. Douglas, 
               JHEP 0305, 46 (2003). arXiv:hep-th/0303194


\bibitem{tH} 
    G. 't Hooft.
     { Nucl. Phys. } {\bf{B357 }}, 221(1991)  .
\bibitem{Page} P. Krtous, D. Kubiznak, D. N. Page, V. P. Frolov,
                JHEP 0702:004,2007;
                V. P. Frolov, D. Kubiznak, Class.Quant.Grav. {\bf{25}}:154005
                ,2008; 
          D. Kubiznak, H. K. Kunduri, Y. Yasui, Phys.Lett. 
{\bf{B678}}, 240 (2009)



                         Zeit. Phys.{\bf{ 65}},589 (1930).
                 W. Pauli and M. Fierz, Hel. Phys. Acta. {\bf{12}}, 297 (1939).
\bibitem{Schild} F. A. E. Pirani and A. Schild, Phys. Rev.{\bf{ 79}}, 
                                           986 (1950);
             F.A. E. Pirani, A. Schild, and Skinner,
          Phys. Rev. {\bf{87}}, 452 (1952).

 \bibitem{Dirac} P. A. M. Dirac, Can. J. Math.{\bf{ 2}}, 129 (1950);
           { \it{ibid}}{\bf{ 3}}, 1 (1951); 
Proc. Roy. Soc. (London) {\bf{A246}}, 326 (1958);
            Phys. Rev.{\bf{ 114}}, 924 (1959).  

\bibitem{Bergmann}P. G. Bergmann, Rev. Mod. Phys. {\bf{21}}, 480 (1949);
                P. G. Bergmann, R. Penfield, R. Schiller, and H. Zatzkis, 
            Phys. Rev.{\bf{ 80}}, 81 (1950);
           P. G. Bergmann, Nuovo Cimento{\bf{ 3}},1177 (1956);
             Helv. Phys. Acta Suppl.{\bf{ 4}}, 79 (1956).



\bibitem{wein2}  S. N. Gupta, Proc. Phys. Soc. {\bf{A65}}, 161 (1952); 
               S. Weinberg, Quantum Theory of Fields, Vol 1, 
                Cambridge University Press, (1996).

\bibitem{Arn} R. Arnowitt, S. Deser, and C. W. Misner, 
                      Phys. Rev. {\bf{117}}, 1595 (1960); 
                                   Phys. Rev. {\bf{120}},313 (1960);
                K. Kucha{\u{r}}, in {\it{Quantum Gravity 2}}, edited by
                  C. J. Isham, R. Penrose, snd D. W. Sciama (Oxford
               Univ. Press, New York, 1981).

\bibitem{Geroch} R. Geroch,
               Phys. Rev.D {\bf{36}}, 1587 (1987).

\bibitem{Sen} A. Sen, J. Math. Phys. {\bf{22}} 1817 (1981 );
                 Phys. Lett. {\bf{119B}} 89 (1982).

\bibitem{Au} G. K. Au, gr-qc/9506001.
 \bibitem{tH2} G. 't Hooft, Phys. Rev. D{\bf{14}}, 3432 (1976).
 





 
\bibitem{sak} J. J. Sakurai, Advanced Quantum Mechanics, Addison Wesley, 
                  Reading, (1967).
\bibitem{bj}  J. D. Bjorken and S. D. Drell, Relativistic Quantum Mechanics,
               McGraw Hill New York (1964).

\bibitem{deWit} B. de Wit and D. Z. Freedman, Phys. Rev. D 
{\bf{21}}, 358 (1980);
J.W. van Holten and  R.H. Rietdijk, hep-th/9205074.
\bibitem{gib}
G.W.Gibbons, R.H. Rietdijk, and J.W. van Holten,  hep-th/9303112,
  { Nucl. Phys. }, {\bf{B404}}, 42 (1993).


\bibitem{Schiff}  L. I. Schiff,  Quantum Mechanics,
               McGraw Hill New York (1968).

\end{thebibliography}
\end{document}